\documentclass[twocolumn,showpacs,preprintnumbers,amsmath,amssymb]{revtex4}

\usepackage{graphicx}
\usepackage{dcolumn}% Align table columns on decimal point
\usepackage{bm}% bold math

\begin{document}

\preprint{APS/123-QED}

%\preprint{?? HEP/123-qed}

\title{Absolute negative conductivity in two-dimensional
electron systems  under microwave radiation} 
\author{ 
Victor~Ryzhii}
%\email{v-ryzhii@u-aizu.ac.jp}
\affiliation{Computer Solid State Physics Laboratory, University of Aizu,
Aizu-Wakamatsu 965-8580, Japan}

\date{\today}

\begin{abstract}
We overview   mechanisms of absolute negative conductivity
in two-dimensional electron systems in a magnetic field irradiated
with microwaves and provide plausible explanations of
the features
observed in recent experiments
related to the so-called zero-resistance (zero-conductance) states.
\end{abstract}



\maketitle
\section{Introduction}

The possibility of absolute negative conductivity (ANC)
when the dc dissipative current is directed opposite to
the local electric field
in two-dimensional electron systems (2DESs)
subjected to a magnetic field and irradiated with microwaves
with the frequency $\Omega$ somewhat exceeding the cyclotron 
frequency $\Omega_c$ or its harmonics
was predicted more than three decades ago~\cite{1} 
(see also Refs.~\cite{2,3,4,5}).
The predicted effect is associated with  photon-assisted
impurity scattering resulting in
electron transitions between the states corresponding to different Landau levels (LLs)
and different positions
of the Larmor orbit centers.  
Two groups of experimentalists observed 
the effect of vanishing
electrical resistance and transition to''zero-resistance''
(in the Hall bar configuration)
and ``zero-conductance'' (in the Corbino samples) states in 2DESs 
under microwave irradiation~\cite{6,7,8,9}.
Similar results were obtained also in Refs.~\cite{10,11}.
Recently, the oscillatory magnetic-field dependences
of the Hall resistivity were observed~\cite{12,13}.
The formation of zero-resistance and zero-conductance
states is primarily attributed to  
photon-assisted impurity scattering mechanism of
ANC~\cite{3,4,5,14} and instability of uniform electric-field
distributions due to  ANC (for example,~\cite{15,16}). 
In this paper, we address to the concept of ANC  invoking
the mechanism associated with the direct  effect of microwave radiation on
the electron scattering processes. 

\section{Phases of  oscillations
and power effects}

The mechanism of ANC associated with
photon-assisted impurity
scattering  explains
 features of the effect. 
According to~\cite{1,2,3,4},
 the positions of
the photoconductivity zeros, maxima, and minima
of the photoconductivity 
correspond mainly to $M\Omega = \Lambda\Omega_c$, 
$M\Omega = \Lambda\Omega_c - \delta^{(+)}\Omega_c$, and 
$M\Omega = \Lambda\Omega_c + \delta^{(-)}\Omega_c$, respectively,
where $\Omega$ and $\Omega_c$ are the microwave and electron cyclotron
frequencies, respectively,
$M,\, \Lambda = 1,2,3,...$ and $\delta^{(\pm)} < 1$
are determined either by the LL broadening $\Gamma$ or by
the net electric field.
Typical calculated dependences of the microwave photoconductivity
corresponding to the abovementioned
positions of the zeros, maxima, and minima are shown in Fig.~1~\cite{4}. 
These dependences correlate well with the experimental curves.
In particular, Fig.~1, shows 
%that  the microwave photoconductivity
%exhibits 
a pronounced zero, maximum, and minimum near two-photon resonance
$2\Omega \simeq 3\Omega_c$.

%%%%%%%%%%%%%%%%%%%%%%%
\begin{figure}[t]
\begin{center}
\includegraphics[width=7.5cm]{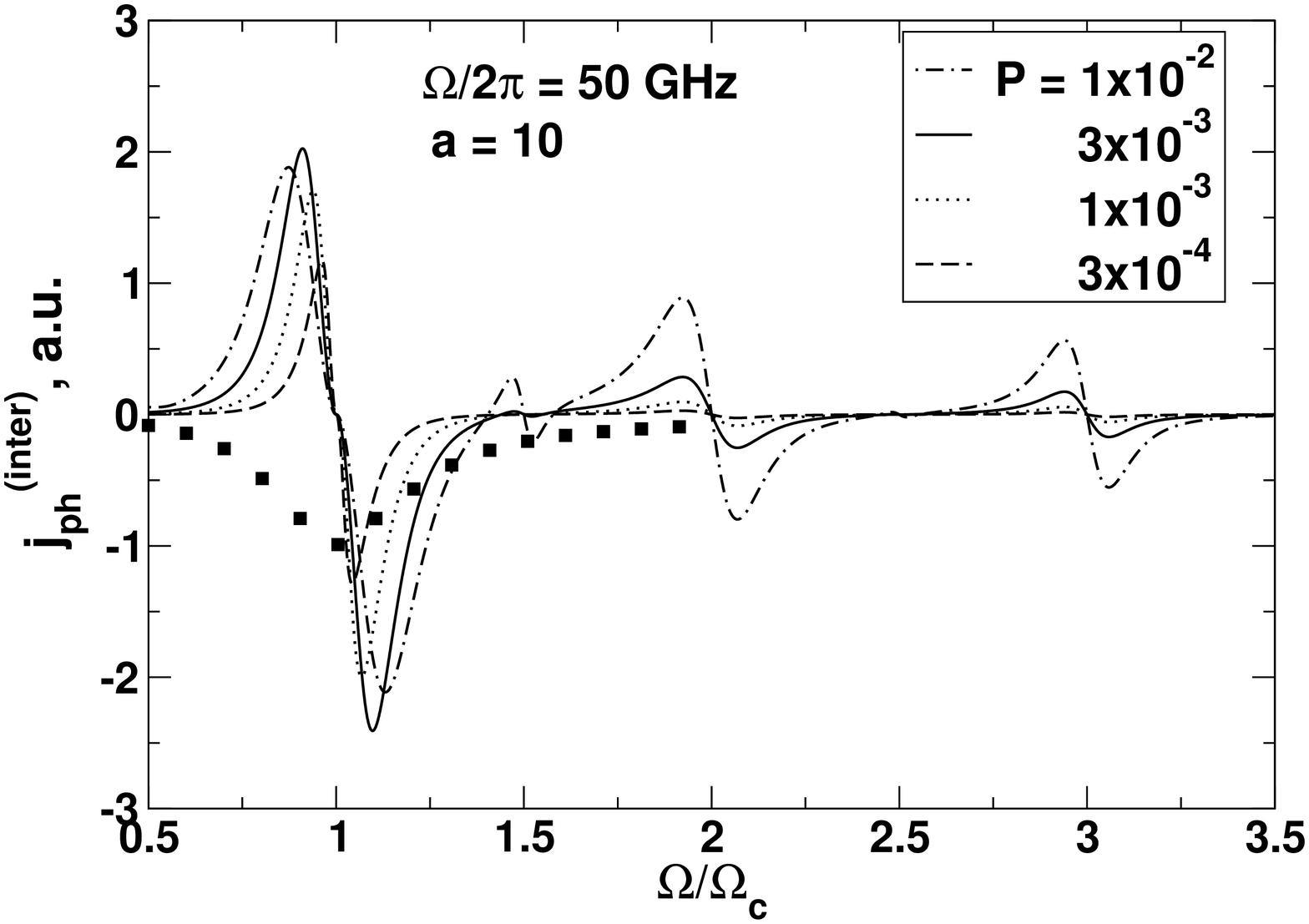}
\end{center}
%\label{fig1}
\caption{Photocurrent due to photon-assisted
impurity scattering
vs $\Omega/\Omega_c$
at different microwave  powers for $a = (\Gamma/\Gamma_i)^2 = 10$
($\Gamma_i$ is the LL broadening due to impurity scattering). 
Markers correspond to contribution
of intra-LL electron transitions.}
\end{figure}

%%%%%%%%%%%%%%%%%%%%%%%
\begin{figure}[t]
\begin{center}
\includegraphics[width=6.7cm]{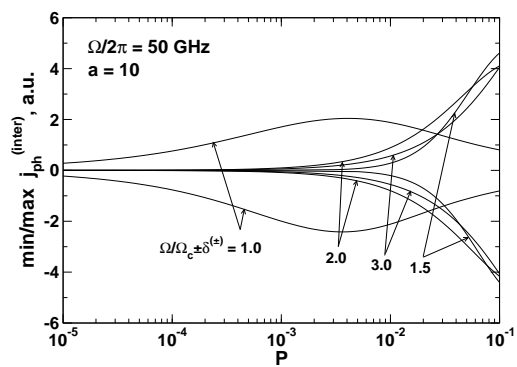}
\end{center}
%\label{fig2}
\caption{Photocurrent maxima and minima corresponding
to different resonances vs normalized microwave power.}
\end{figure}

As seen from Fig.~1,
the magnitude of the first maximum and minimum ($\Lambda = 1$)
as a function of microwave power tends to saturation
when this power increases,  while the span of the maxima and minima
corresponding to higher resonances (see Fig.~2)
continues to increase with  increasing
microwave power ~\cite{4}. 
%This leads to 
%relatively large magnitude
%of these maxima and minima~\cite{4}.

\section{Temperature suppression of ANC}

%%%%%%%%%%%%%%%%%%%%%%%
\begin{figure}[t]
\begin{center}
\includegraphics[width=6.5cm]{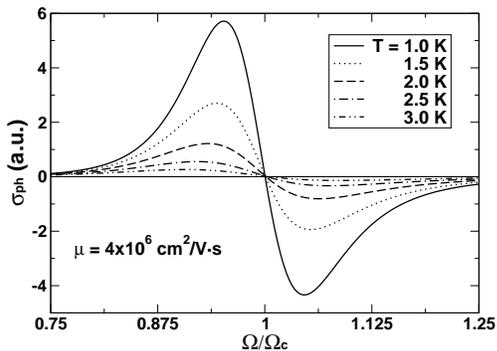}
\end{center}
%\label{fig2}
\caption{Photoconductivity associated with  photon-assisted
impurity scattering vs $\Omega/\Omega_c$
at different temperatures.}
\end{figure} 
%%%%%%%%%%%%%%%%%%%%%%%
\begin{figure}[t]
\begin{center}
\includegraphics[width=6.5cm]{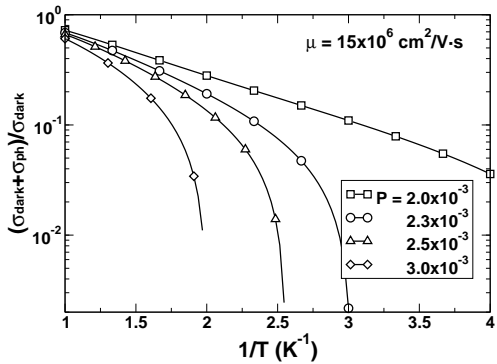}
\end{center}
%\label{fig2}
\caption{Temperature dependences of normalized dissipative conductivity   
corresponding to  different microwave powers.}
\end{figure} 

A significant
sensitivity of the amplitude
of  microwave photoconductivity oscillations to the temperature
can be attributed to an increase of the LL broadening 
due to increasing dependence  of the electron-electron 
scattering. This is because the magnitude of the maxima and minima
sharply decreases with increasing  LL broadening.
The  microwave photoconductivity is actually determined by
the processes of both absorption and emission of photons.
This results in an additional temperature dependent factor:
$\sigma_{ph} \propto [1 - \exp(-\hbar\Omega/T)]$. At the photon energies
$\hbar\Omega$ corresponding to the experimental values, this factor
markedly decreases when the temperature increases from $T \lesssim 1$~K
to $T \simeq 3- 5$~K. Considering this and 
assuming that the contribution of  the electron-electron interaction
to the  LL broadening 
$\Gamma_e \propto T^2\ln (\varepsilon_F/T)$~\cite{17,18},
where $T$  and $\varepsilon_F$ are the temperature and electron Fermi energy 
($\varepsilon_F \gg \hbar\Omega_c, T$), respectively,
one can find that the shape of the first maximum and minimum
transforms with temperature as shown in Fig.~3. Fig.~4
demonstrates the temperature dependences of the net dissipative
conductivity at different microwave powers.
%%%%%%%%%%%%%%%%%%%%%%%
\begin{figure}[t]
\begin{center}
\includegraphics[width=6.5cm]{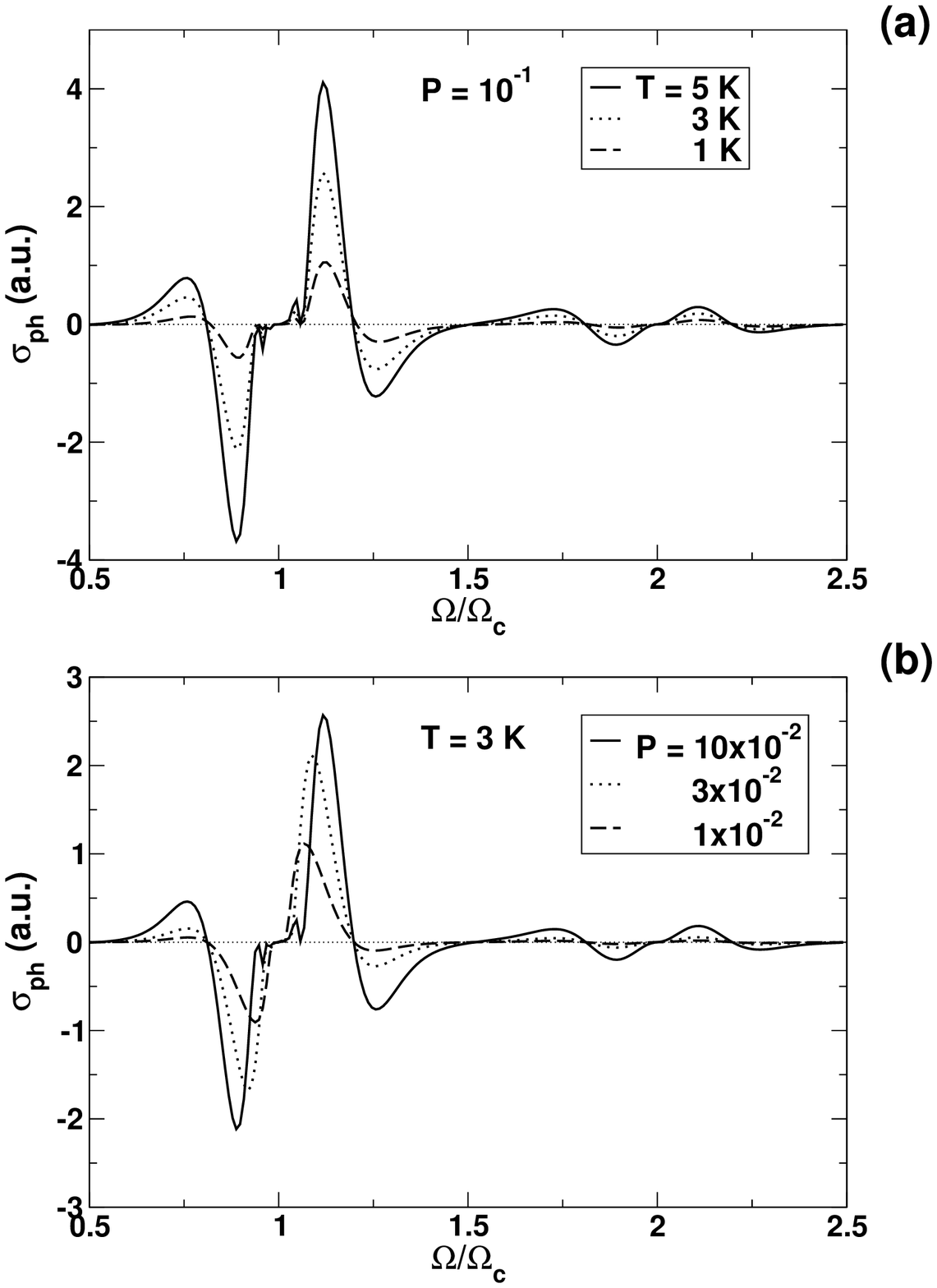}
\end{center}
%\label{fig2}
\caption{Contribution of photon-assisted scattering
on acoustic phonons to photoconductivity as a function of $\Omega/\Omega_c$
at different (a) temperatures and (b)  microwave  powers.}
\end{figure} 
Photon-assisted scattering of electrons on acoustic phonons
can also provide  an  oscillatory contribution to 
the microwave photoconductivity~\cite{19,20}. 
The pertinent dependences calculated in~\cite{19} are shown in Fig.~5.
As seen from Fig.~5,  photon-assisted acoustic scattering
leads to oscillatory microwave photoconductivity
with the phase virtually opposite
to the phase of the oscillation associated with  photon-assisted impurity
scattering. Since the contribution of the ``phonon'' mechanism
increases with temperature, this mechanism can 
be weak at, for instance,  $T \lesssim 1$~K
and 
be essential
 at elevated temperatures ($T \simeq 3- 5$~K).
The contribution of  photon-assisted acoustic scattering
to the microwave photoconductivity can be marked in 2DESs
with rather high electron mobility, i.e., in the samples
with sufficiently weak impurity scattering.
As a result, the mechanism of ANC associated with 
 photon-assisted acoustic scattering 
in 2DESs with very large
mobilities can effectively interfere with
 the photon-assisted impurity
scattering mechanism  and surpass it even at fairly low temperatures.
%The acoustic~\cite{21} and photon-assisted acoustic scattering mechanisms
%can also yield some contribution to the LL broadening and, then,
%to a smearing of the microwave photoconductivity
%oscillations and vanishing of ANC. 

\section{Influence of microwaves on the Hall effect}
%%%%%%%%%%%%%%%%%%%%%%%
\begin{figure}[t]
\begin{center}
\includegraphics[width=6.5cm]{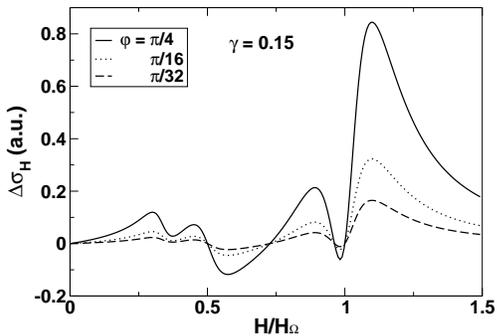}
\end{center}
%\label{fig2}
\caption{The Hall microwave photoconductivity
vs normalized magnetic field 
($\gamma = \Gamma/\Omega_c = 0.15$).}
\end{figure}

The photon-assisted impurity scattering provides contributions
to both the dissipative and Hall components of the conductivity
(and, consequently, resistivity) tensor. Fig.~6 shows
the variation of the Hall conductivity under microwave radiation
 vs magnetic field
normalized by $H_{\Omega} \propto \Omega$. It is instructive that
the Hall microwave photoconductivity is an even function
of the magnetic field. This component is sensitive to
the polarization of microwave radiation. In the case of linear
polarization, $\Delta\sigma_H \propto \sin2\varphi$~\cite{21},
as predicted a long time ago~\cite{22} (see also,~\cite{5}), where $\varphi$
is  the angle between the dc electric field and the ac microwave field.
The dissipative and Hall microwave photoconductivities affect
both the dissipative and Hall resistivities. Fig.~7
shows the calculated  dissipative resistivity and variation
of the Hall resistivity as functions of the normalized magnetic
field~\cite{22}. These dependences are similar to
the experimental
one's~\cite{12,13}. 
%%%%%%%%%%%%%%%%%%%%%%%
\begin{figure}[t]
\begin{center}
\includegraphics[width=6.5cm]{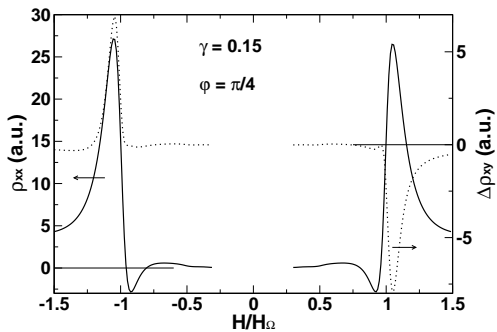}
\end{center}
%\label{fig2}
\caption{Dissipative (solid curves) and Hall (dotted curves)
resistivities vs normalized magnetic field.}
\end{figure} 

\section{Comments}

Apart from the mechanisms of ANC
associated with the effect of the microwave field
on the electron dynamics and  scattering
processes, the effect of ANC in a 2DES
subjected to  a magnetic field
can also be associated with a deviation of the electron
distribution function from equilibrium one (``statistical'' mechanism). 
A model based on the assumption that the electron distribution
function is nonmonotonic with inversion population of the
states near the center of LL's caused by the absorption
of microwave radiation was  proposed~\cite{11,23}. 
In this model, a photoexcited electron contributes to
the  dissipative 
photoconductivity immediately when it  transfers to an upper LL
absorbing a photon (this actually corresponds to the dynamic
mechanism) and  owing to non-radiative scattering events upon the excitation.
The impurity scattering of the photoexcited
electrons can  provide a negative contribution
to the  dissipative 
photoconductivity if
 the maxima of the electron density of states
are shifted with respect to the maxima of the distribution 
function~\cite{24}.   
The main problem is, however, the feasibility of such nontrivial
electron distributions, particularly, due to both intra-LL and inter-LL
electron-electron scattering~\cite{18,25} 
and scattering of electrons on  acoustic phonons~\cite{26}. 
The dynamical and statistical
mechanisms can be distinguished by their different polarization
selectivity.

In summary, we demonstrated that the main features
of the mechanism of ANC associated with  photon-assisted
scattering of electrons on impurities and acoustic phonons
are consistent with the results of experimental observations.

The author is grateful to R.~Suris, V.~Volkov, A.~Chaplik,
and V.~Vyurkov for
discussions and to I.~Khmyrova and A.~Satou
for assistance.

\end{document}